\documentclass[aps,prx,superscriptaddress,twocolumn,showpacs,amsmath,amssymb,reprint,longbibliography,floatfix]{revtex4-2}

\usepackage{graphicx}
\usepackage{tikz}
\usepackage[percent]{overpic}
\usepackage{placeins}
\usepackage{dcolumn}
\usepackage{bm}
\usepackage[T1]{fontenc}
\usepackage[utf8]{inputenc}
\usepackage{xcolor}
\usepackage{booktabs}
\usepackage[colorlinks=true,allcolors=black,citecolor=blue]{hyperref}

\usepackage{amsfonts}
\usepackage{mathptmx}
\begin{document}
\raggedbottom

\title{Knowledge Transfer from a Universal Interatomic Potential to a Universal Electron Density Model}

\author{Xin Chen}
\email{xin.chen@iat.cn}
\affiliation{Samuel C. C. Ting Laboratory, Shandong Institute of Advanced Technology, Jinan 250100, Shandong Province, People's Republic of China}

\author{Xinyan Zhang}
\affiliation{College of Computer Science and Technology, Harbin Engineering University, Harbin 150001, China}

\author{Duo Wang}
\email{duo.wang@mpu.edu.mo}
\affiliation{Faculty of Applied Sciences, Macao Polytechnic University, Macao SAR, 999078, People's Republic of China}

\date{\today}

\begin{abstract}
Energies and forces are available for far more atomic configurations than are reference electron densities. Can representations learned by a universal interatomic potential from this broader range of environments help a density model predict both static charge densities and where charge moves under perturbation? We connect a frozen EqV3 OMat-G encoder to the EdenGNN-Uni electron-density decoder through a dual-parity $O(3)$-equivariant bridge, the Potential--Density Bridge (PxD-Bridge). EdenGNN-Uni supplies all teacher hidden states and density targets on 50,000 bulk structures, so transfer requires no new density-functional-theory density labels. Evaluated without calibration, the PxD-Bridge model gives the lower mean total-density error on 965 held-out bulk crystals, on all four held-out MoS$_2$ nanotubes, and for both the unstrained and strained structures in the 100-crystal test. The decisive test is the density response: normalized by the total valence-electron count, the mean strain-induced charge-redistribution error decreases from 0.739$\%$ for EdenGNN-Uni to 0.594$\%$ for PxD-Bridge, with the PxD-Bridge model more accurate on 85 of the 100 crystals. Across an expanded 579-crystal benchmark spanning 79 elements under lattice strain and atomic rattling, the mean redistribution error decreases from 0.467$\%$ to 0.384$\%$, with PxD-Bridge more accurate on 543 of the 579 crystals. In a matched control using the identical bridge architecture, randomly initialized EqV3 fails to surpass EdenGNN-Uni, whereas pretrained EqV3 gives a lower redistribution error than random EqV3 on 95 of the 100 strained crystals and 549 of the 579 full-periodic-table crystals. Thus, energy-and-force pretraining can change how a distilled density model generalizes across structures and perturbations, allowing a student trained only to imitate its teacher to exceed that teacher on both static densities and electronic response.
\end{abstract}

\maketitle

\section{Introduction}
A central challenge in scientific machine learning is to reuse information learned from abundant, inexpensive supervision to predict physical observables that are expensive to label. In first-principles materials data, total energies and atomic forces are compact labels that are routinely retained from high-throughput density-functional theory (DFT) workflows, whereas electron densities are high-dimensional fields whose storage, curation, and supervision are substantially more demanding. This imbalance has produced a much broader data foundation for universal interatomic potentials~\cite{pbM3G,pbCHG,pbSeven,pbDPA,pbMatter,pbMACEMP,pbMACEOMAT,pbMACEMatPES,pbMACEMH,pbMagMACE,pbField,pbPolar,pbEqV3,pbEqV3Models,pbPETMAD,pbGRACE,pbNEP89} than for universal electron-density models~\cite{pbChargE3,pbEAC,pbEden}: OMat24 contains more than one hundred million DFT calculations suitable for learning energies and forces~\cite{pbOMat}, while EdenGNN-Uni was trained on charge densities for approximately fifty thousand bulk structures~\cite{pbEden}. This disparity raises a broader transfer-learning question: can a model trained on abundant scalar and vector supervision acquire representations that are useful for predicting a scarce field observable? For electron densities, the relevant test is not only whether the static total density can be reproduced, but whether the model accurately captures how charge redistributes when atomic geometry is perturbed.

A direct way to use an interatomic potential would be to pass its predicted energy and forces to a density model. This is unlikely to be sufficient. The same vanishing force condition can occur in many equilibrium structures with very different bond directions, orbital hybridizations, and charge anisotropies. Electron densities therefore depend on geometric information that is only indirectly constrained by scalar energies and vector forces. The more useful object is the potential's internal representation: a set of local atomic features that encode chemical environment and directional response before these features are compressed into energy and force outputs~\cite{pbMACEArch}. If these features can be placed in the representation space expected by a density decoder, they may provide a route for transferring information from abundant energy-and-force supervision to scarce density-field supervision.

This transfer problem is also a model-interface problem. Recent work on frozen neural networks has shown that learned interfaces can connect independently trained models~\cite{pbStitch} by querying or transforming hidden states, as in XBridge~\cite{pbXBridge}, LatBridge~\cite{pbLatBridge}, and Latent Bridge~\cite{pbLatentBridge}. Atomistic models add a geometric constraint that is absent from most language-model interfaces: their features carry $O(3)$ rotational and parity structure~\cite{pbLewis}. For two models evaluated on the same crystal, the atomic sites are already registered, but scalar, vector, pseudovector, and higher-order tensor components must transform consistently under rotations and spatial inversion. The bridge between a potential and a density model should therefore be not only learned, but $O(3)$-equivariant and parity-complete.

A further complication is that density labels need not be used to train the bridge. If the density model supplies all hidden-state, expansion-coefficient, and density-field targets, the student is trained to reproduce its teacher rather than DFT. On the training distribution, exact reproduction would simply inherit the teacher's errors. The decisive test is therefore generalization to held-out crystals, non-bulk geometries, and finite perturbations: can a student driven by a pretrained interatomic potential move closer to the DFT density than the teacher that supplied its sole supervision? This teacher-supervised setting makes any systematic improvement over the teacher especially informative, because it cannot be attributed to additional DFT density labels.

In this work, we introduce the Potential--Density Bridge (PxD-Bridge) framework by connecting the frozen EqV3 OMat-G universal interatomic potential~\cite{pbEqV3} to the decoder of EdenGNN-Uni through a dual-parity $O(3)$-equivariant bridge. We name each implementation by its potential encoder and density decoder, in that order, and refer to PxD-Bridge (EqV3 OMat-G--EdenGNN-Uni) as the PxD-Bridge model. At each atom, the bridge reads per-degree normalized multi-layer representations (blocks 2, 4, and 7) from frozen EqV3, retains both natural-parity ($X^+$) and unnatural-parity ($X^-$) components from the same two inversion-paired forward passes, and maps them together with short-range bond moments into the 842-component hidden state expected by the EdenGNN-Uni decoder. EdenGNN-Uni supplies all training targets on its original 50,000 bulk structures, with no new DFT density labels entering training or checkpoint selection.

The results demonstrate that representation transfer from an interatomic potential improves both static density accuracy and physical response under perturbation. Evaluated without calibration across all locked DFT benchmarks, PxD-Bridge surpasses its EdenGNN-Uni teacher: mean total-density error decreases on 965 held-out bulk crystals ($1.2407\%$ versus $1.2463\%$) and on all four held-out MoS$_2$ nanotubes ($0.9849\%$ versus $1.0408\%$). On 100 separate crystals paired with random 2--10\% uniaxial or biaxial strains, PxD-Bridge lowers mean total-density error both before strain ($1.048\%$ versus $1.143\%$, lower on 70 crystals) and after strain ($1.234\%$ versus $1.402\%$, lower on 80 crystals). For the strain-induced charge redistribution normalized by the total valence-electron count, the mean error decreases from $0.739\%$ for EdenGNN-Uni to $0.594\%$ for PxD-Bridge---a $19.7\%$ relative reduction, with lower error on 85 of the 100 crystals and statistically significant gains under both tension and compression. Extending the evaluation to 579 held-out crystals spanning 79 elements under both lattice strain (290 pairs) and fixed-cell atomic rattling (289 pairs), PxD-Bridge outperforms EdenGNN-Uni on charge redistribution in 543 of 579 crystals ($93.8\%$, including 280/289 under atomic rattling) and achieves a $>50\%$ win rate across all 79 tested elements. Controlled ablations isolate both the contribution of the unnatural-parity EqV3 channel ($X^-$) and the essential role of energy-and-force pretraining: substituting a frozen randomly initialized EqV3 encoder into the identical dual-parity bridge architecture fails to surpass EdenGNN-Uni, whereas pretrained EqV3 improves over the random encoder on 95/100 strained pairs and 549/579 full-periodic-table crystals ($94.8\%$).

\section{Model and Methods}

The interatomic potential and the density model are evaluated on the same periodic structure: atomic species $Z_i$, Cartesian positions $\mathbf R_i$, and lattice matrix $C$. The atom ordering is therefore shared, so atom $i$ in EqV3 OMat-G corresponds directly to atom $i$ in EdenGNN-Uni. This common indexing removes the need for a learned atom-matching step, and the bridge only has to align the $O(3)$-equivariant representations attached to corresponding atomic sites.

\subsection{Model interfaces and dual-parity multi-layer readout}

\begin{figure}[!t]
\centering
\includegraphics[width=8.3cm]{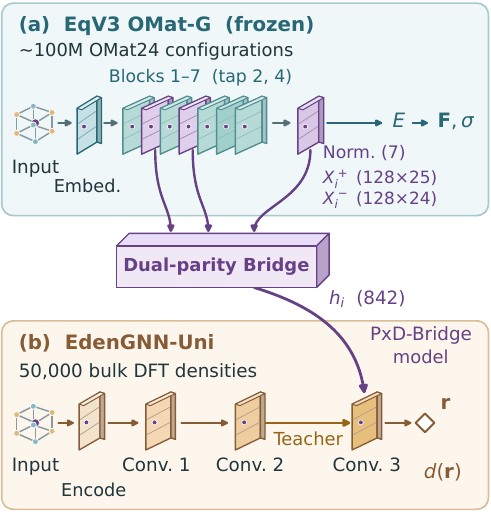}
\caption{Where the interatomic potential connects to the density model. Purple dots track the same atomic site $i$ across layers in both networks. (a) Frozen EqV3 OMat-G is tapped at Transformer blocks 2 and 4 and after final normalization (block 7), supplying the dual-parity bridge with natural-parity ($X_i^+$, $128\times25$) and unnatural-parity ($X_i^-$, $128\times24$) atomic representations from right-handed inversion-paired passes. The native energy readout yields $E$, whose derivatives give forces and stress~\cite{pbEqV3}. (b) The gold arrow carries the output of convolution 2 to convolution 3 in the original EdenGNN-Uni teacher~\cite{pbEden}. In the PxD-Bridge model, the purple arrow carries the 842-component $O(3)$ hidden state $h_i$ to convolution 3. The decoder evaluates the deformation density $d(\mathbf r)$ at the probe marked by the diamond.}\label{fig:model-interfaces}
\end{figure}

Figure~\ref{fig:model-interfaces} defines the interface used throughout this work. On the potential side [Fig.~\ref{fig:model-interfaces}(a)], EqV3 OMat-G processes a 6~\AA{} neighbor graph with seven Transformer blocks, each containing eight attention heads~\cite{pbEqV3}. We freeze the entire EqV3 OMat-G encoder and tap three depths along the network: the outputs of blocks 2 and 4 and the final normalized representation after block 7, before the energy readout. At each tapped layer $b\in\{2,4,7\}$, the raw atomic tensor has 128 channels over angular degrees $\ell=0,\ldots,4$ (25 magnetic components per channel), evaluated in the cyclic $(y,z,x)$ coordinate convention used at the EdenGNN-Uni interface.

Because EqV3 OMat-G is constructed to be $SO(3)$-equivariant rather than explicitly parity-separated, evaluating it on a crystal $\mathcal S=(\mathbf R,C)$ and its spatial inversion $\bar{\mathcal S}=(-\mathbf R,C)$---where negating positions while retaining the right-handed lattice matrix $C$ ($\det C>0$) preserves periodic cell orientation and neighbor-list conventions---allows exact decomposition into definite inversion parities at zero extra encoder cost relative to standard parity symmetrization. Denoting the layer-$b$ outputs for the original and inverted geometries by $A_i^{(b)}$ and $\bar A_i^{(b)}$, we form both the natural-parity tensor $X_i^{+(b)}$ [parity $(-1)^\ell$] and the unnatural-parity tensor $X_i^{-(b)}$ [parity $-(-1)^\ell$]:
\begin{align}
 X_{i,\ell m}^{+(b)} &= \tfrac{1}{2}\!\left[A_{i,\ell m}^{(b)} + (-1)^\ell \bar A_{i,\ell m}^{(b)}\right] \quad (\ell=0,\ldots,4), \\
 X_{i,\ell m}^{-(b)} &= \tfrac{1}{2}\!\left[A_{i,\ell m}^{(b)} - (-1)^\ell \bar A_{i,\ell m}^{(b)}\right] \quad (\ell=1,\ldots,4).
\end{align}
Thus $X_i^{+(b)}$ has shape $128\times25$ and $X_i^{-(b)}$ has shape $128\times24$ (omitting the $\ell=0$ pseudoscalar, which vanishes in the downstream target). Retaining $X_i^{-(b)}$ preserves the encoder's native pseudovector and pseudotensor response and routes it directly into the `1e`, `2o`, `3e`, and `4o` channels rather than relying exclusively on tensor products of natural-parity features.

Raw feature magnitudes vary by roughly an order of magnitude across EqV3 blocks 2, 4, and 7 and across angular degrees $\ell$. Before combining layers, we divide each block and degree by an $SO(3)$-invariant ($m$-independent) root-mean-square scale calibrated on training crystals, $\tilde X_i^{\pm(b)}=X_i^{\pm(b)}/\sigma^{\pm(b)}$. Full-channel ($128\to128$) residual mixing anchored at block 7 then produces unified natural- and unnatural-parity representations:
\begin{equation}
 \label{eq:layer-readout}
 f_{i,\ell}^{\pm} = \tilde X_{i,\ell}^{\pm(7)} + W_{2,\ell}^{\pm}\tilde X_{i,\ell}^{\pm(2)} + W_{4,\ell}^{\pm}\tilde X_{i,\ell}^{\pm(4)},
\end{equation}
with zero-initialized mixing matrices $W_{2,\ell}^{\pm}, W_{4,\ell}^{\pm}\in\mathbb R^{128\times128}$ acting identically on all magnetic components $m$ within degree $\ell$.

On the density side [Fig.~\ref{fig:model-interfaces}(b)], EdenGNN-Uni predicts the deformation density using a sequence of atomic convolutions followed by a probe-position density decoder~\cite{pbEden}. We place the bridge between the second and third atomic convolutions. The bridge produces, for each atom, an EdenGNN-Uni-compatible hidden state $h_i$ with the same 842-component channel ordering and $O(3)$ irrep structure as the teacher state passed to convolution 3. The subsequent EdenGNN-Uni decoder is copied into the PxD-Bridge model and evaluates the deformation density at arbitrary probe positions. Adding the superposed isolated-atom densities gives the total valence density.

\subsection{Dual-parity equivariant bridge and density readout}

\begin{figure*}[!t]
\centering
\includegraphics[width=17.78cm]{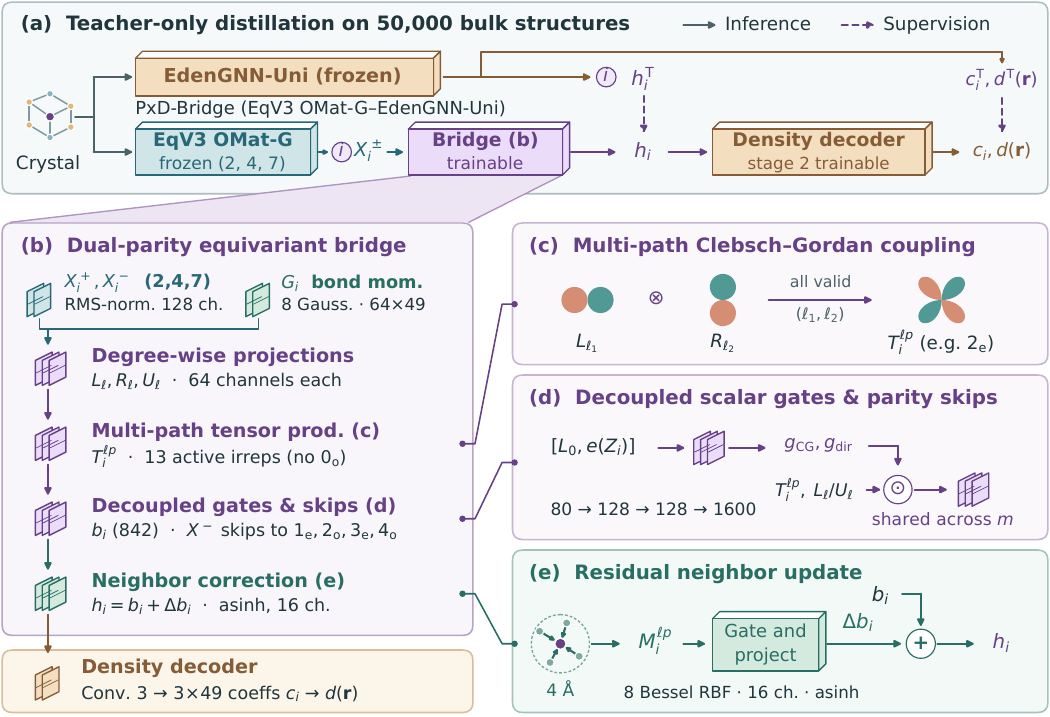}
\caption{Teacher-only distillation and the dual-parity $O(3)$-equivariant bridge. (a) On each of 50,000 bulk structures, solid arrows trace inference from frozen EqV3 OMat-G (blocks 2, 4, and 7) through the trainable bridge and student decoder. Frozen EdenGNN-Uni supplies all hidden-state ($h_i^{\rm T}$), expansion-coefficient ($c_i^{\rm T}$), and deformation-density [$d^{\rm T}(\mathbf r)$] targets (dashed purple arrows). (b) The 64-channel dual-parity bridge combines RMS-normalized $X_i^+, X_i^-$ with 8-Gaussian bond moments $G_i$ [Eq.~(\ref{eq:bridge-moments})], applies multi-path Clebsch--Gordan couplings and decoupled dual-parity skips to form $b_i$, and adds a 16-channel neighbor correction to give $h_i=b_i+\Delta b_i$, which enters the gold density decoder [Eq.~(\ref{eq:density-probe})]. (c) Fixed Clebsch--Gordan contractions couple all valid degree pairs $(\ell_1,\ell_2)$ into each of the 13 active irreps. (d) A scalar context MLP ($80\to128\to128\to1600$) generates separate channel gates for tensor-product paths and direct natural/unnatural-parity skips, broadcast across $m$. (e) Within 4~\AA{}, neighbor messages built from 8 Bessel radial basis functions and non-saturating $\mathrm{asinh}$ scalar features [Eq.~(\ref{eq:bridge-message})] are gated and projected to form $\Delta b_i$.}\label{fig:physbridge}
\end{figure*}

The solid path in Fig.~\ref{fig:physbridge}(a) follows one prediction from atomic features to electron density. When a structure $\mathcal S$ is transformed by $Q\in O(3)$ and translated by $\mathbf t$, its density obeys $\rho_{Q\mathcal S+\mathbf t}(Q\mathbf r+\mathbf t)=\rho_{\mathcal S}(\mathbf r)$ at the corresponding probe. The bridge enforces exact $O(3)$ equivariance by combining angular representations exclusively through degree-preserving linear maps, $SO(3)$-invariant scalar gates, spherical harmonics, and fixed Clebsch--Gordan couplings. A sitewise dual-parity tensor module first combines $(f_i^+,f_i^-)$ with local bond moments $G_i$ to form a base state $b_i$; a non-saturating residual neighbor update then yields $h_i=b_i+\Delta b_i$. Figure~\ref{fig:physbridge}(b) traces these stages, and panels (c)--(e) resolve the multi-path tensor product, decoupled gates, and neighbor update.

\paragraph{Short-range learned bond moments.} EqV3 supplies angular degrees up to $\ell=4$, whereas EdenGNN-Uni's hidden state extends to $\ell=6$ across both parities with multiplicities $(64,20,16,8,6,5,4)$ for $\ell=0,\ldots,6$ (842 total entries per atom). To supply short-range geometric resolution and $\ell=5,6$ angular channels within $r_c=4.0$~\AA{}, we compute degree-normalized local bond moments:
\begin{equation}
 \label{eq:bridge-moments}
 G_{i,as\ell m} = \frac{1}{\sigma_{a\ell}\sqrt{1+\sum_{k\in\mathcal N_i} f_{ik}}}\sum_{j\in\mathcal N_i} B_a(r_{ij})\,s_{js}\,Y_{\ell m}(\hat{\mathbf r}_{ij}),
\end{equation}
where $f_{ij}=[1+\cos(\pi r_{ij}/r_c)]/2$ for $r_{ij}<r_c$ is a smooth cosine envelope, $B_a(r_{ij})=f_{ij}\exp[-(r_{ij}-\mu_a)^2/(2\sigma_r^2)]$ uses eight evenly spaced Gaussian centers $\mu_a\in[0.75,3.75]$~\AA{} ($\sigma_r=0.35$~\AA{}) providing fine short-bond radial coverage, and $s_{js}$ is entry $s$ of an eight-component neighbor scalar vector comprising unity, $Z_j/35$, and six learned LayerNorm-normalized linear projections of $f_{j,0}^+$. Division by the calibrated unit-source RMS scale $\sigma_{a\ell}$ standardizes each $(a,\ell)$ block, yielding $64\times49$ natural-parity geometric features spanning $\ell=0,\ldots,6$.

\paragraph{Multi-path Clebsch--Gordan couplings and dual-parity skips.} For $\ell\le4$ we concatenate $f_{i,\ell}^+$ and $G_{i,\ell}$ ($128+64=192$ channels); for $\ell=5,6$ the natural-parity input comes from $G_{i,\ell}$ alone ($64$ channels). After an $SO(3)$-equivariant soft norm guard $x\mapsto x[1+\langle x^2\rangle_m/20^2]^{-1/2}$, degree-wise bias-free linear projections map the inputs to width $w=64$: left factors $L_\ell$ ($\ell=0,\ldots,6$), right factors $R_\ell$ ($\ell=1,\ldots,4$), and unnatural-parity projections $U_\ell$ ($\ell=1,\ldots,4$) directly from $f_{i,\ell}^-$.

In EdenGNN-Uni's hidden state, the `64x0o` pseudoscalar block is identically zero for all crystals; we omit it and operate on the 13 active degree--parity irreps. Every active irrep $(\ell,p)$ combines all valid $(\ell_1,\ell_2)$ couplings among $(L_{\ell_1},R_{\ell_2})$ [Fig.~\ref{fig:physbridge}(c)]: e.g., $(1,1),(2,2),(3,3),(4,4)$ for $0_{\rm e}$ and $1_{\rm e}$; $(1,1),(2,2),(1,3),(3,3),(2,4)$ for $2_{\rm e}$; $(1,2),(2,3),(1,4),(3,4)$ for $3_{\rm o}$; and $(3,3),(2,4),(4,4)$ for $5_{\rm e}$ and $6_{\rm e}$. Each pair is contracted channel by channel via Wigner-$3j$ coefficients scaled by $\sqrt{2\ell+1}$, gated, concatenated, and linearly projected to the decoder multiplicity.

A two-hidden-layer SiLU context network ($80\to128\to128\to1600$) takes $[L_{i,0},e(Z_i)]$, where $e(Z_i)\in\mathbb R^{16}$ is a zero-initialized element embedding (providing safe fallback for unseen elements), and produces 64 even-scalar shifts together with $(13+11)\times64=1536$ decoupled bounded gates $g=1+2\tanh(\hat g/2)$ [Fig.~\ref{fig:physbridge}(d)]. Separating tensor-product gates ($13\times64$) from direct-skip gates ($11\times64$) allows the network to weight direct skips independently of higher-order couplings: all seven natural-parity groups ($\ell=0,\ldots,6$) receive gated linear skips from $L_\ell$, while the four unnatural-parity groups (`1e`, `2o`, `3e`, `4o`) receive gated linear skips directly from $U_\ell$. Learnable per-group log-scales (clamped to $[-2,2]$), calibrated target RMS scales, and per-element scalar centers assemble the 842-component base state $b_i$.

\paragraph{Non-saturating residual neighbor update.} Figure~\ref{fig:physbridge}(e) illustrates the 4~\AA{} residual refinement $\Delta b_i$. Scalar features are first standardized relative to per-element centers and transformed with a non-saturating inverse hyperbolic sine, $\tilde s_i = 3\,\mathrm{asinh}[(b_{i,0}-c_{Z_i})/(3\sigma_0)]$. Edge distances within $r_c=4.0$~\AA{} are encoded by eight orthogonal sinusoidal Bessel radial basis functions $\phi_k(r_{ij})=f_{ij}\sin(k\pi r_{ij}/r_c)$ ($k=1,\ldots,8$) mapped by a two-layer MLP to $C=16$ radial filter channels $R(r_{ij})$. For each of the 13 active irreps, neighbor features are projected to 16 channels; on natural-parity irreps, a learned scalar-context amplitude multiplies $Y_{\ell m}(\hat{\mathbf r}_{ij})$ and adds directional edge information. Messages are aggregated as
\begin{equation}
 \label{eq:bridge-message}
 M_i^{\ell p} = \sum_{j\in\mathcal N_i} w_{ij}\,R(r_{ij})\odot v_{ij}^{\ell p},
\end{equation}
with $w_{ij}=f_{ij}/(1+\sum_{k\in\mathcal N_i} f_{ik})$, gated per channel by bounded scalar gates at receiving atom $i$, and linearly projected back to the decoder channel multiplicities via zero-initialized weights, yielding $h_i=b_i+\Delta b_i$.

\paragraph{From hidden states to density.} The student decoder (convolution 3 and probe head copied from EdenGNN-Uni) maps $h_i$ to three sets of 49 angular coefficients $c_{ik\ell m}$ per atom and evaluates the deformation density at any probe $\mathbf r$:
\begin{equation}
 \label{eq:density-probe}
 d(\mathbf r) = \sum_{ik\ell m} c_{ik\ell m}\,F_{k\ell}(Z_i,r_i)\,Y_{\ell m}(\hat{\mathbf u}_i),
\end{equation}
where $\mathbf u_i=\mathbf r-\mathbf R_i$, $r_i=|\mathbf u_i|$, $F_{k\ell}$ is a learned species-dependent radial weight, and the sum runs over contributing periodic images. Adding the superposition of isolated-atom pseudo-densities (SAD) yields the total valence density $\rho(\mathbf r)$ compared with DFT.

\subsection{Teacher-only distillation and controlled ablations}

The dashed paths in Fig.~\ref{fig:physbridge}(a) specify the teacher supervision. We reuse the exact 50,000 bulk crystals from EdenGNN-Uni's training set and run the frozen teacher on each structure to obtain its second-convolution hidden state $h_i^{\rm T}$, output expansion coefficients $c_i^{\rm T}$, and deformation-density field $d^{\rm T}(\mathbf r)$. No DFT density enters the transfer loss or checkpoint selection.

Each training step evaluates the student and teacher at 256 randomly sampled probes per crystal and minimizes
\begin{equation}
 \mathcal L = \mathcal L_{\rm field} + w_h \mathcal L_{\rm hidden} + w_c \mathcal L_{\rm coeff},
\end{equation}
where $\mathcal L_{\rm field}$ is the per-structure mean-squared field error divided by the teacher's mean-squared field (floored at $10^{-6}$), $\mathcal L_{\rm hidden}$ is the standardized squared hidden-state error averaged equally across the 13 active irrep groups, and $\mathcal L_{\rm coeff}$ is the RMS-standardized squared error against the noiseless teacher coefficients $c_i^{\rm T}$.

Training follows a two-stage cosine schedule with AdamW (weight decay $10^{-4}$, gradient norm clip $1.0$, seed $922806$, batch size 2 packs): Stage~1 trains only the bridge for 4 epochs with the decoder frozen (learning rate $3\times10^{-4}\to3\times10^{-5}$, $w_h=w_c=0.05$); Stage~2 jointly optimizes the bridge ($1.5\times10^{-4}\to10^{-6}$) and student decoder ($2\times10^{-5}\to10^{-6}$) for 6 epochs ($w_h=w_c=0.02$). Checkpoints are selected strictly by lowest pooled relative RMSE against the frozen teacher ($R_{\rm val}$, defined below) at 512 fixed probes per structure across 1,000 separate official validation crystals.

PxD-Bridge contains 764,656 trainable bridge parameters and 2,065,295 decoder parameters (2,829,951 total). On the 1,000-crystal teacher validation set, $R_{\rm val}$ decreases monotonically across all 10 epochs and reaches $5.6368\%$. To isolate the contribution of the unnatural-parity EqV3 channel $X_i^-$, we train a matched natural-parity ablation (without $X^-$, 564,592 bridge parameters, 2,629,887 total) with identical data, seed, normalization, and schedule; its validation error stops at $5.7978\%$, confirming that retaining unnatural-parity EqV3 features improves teacher reproduction by 0.161 percentage points at zero extra encoder cost.

To isolate the contribution of energy-and-force pretraining in the frozen potential backbone itself, we construct a matched random-encoder control directly on the full dual-parity PxD-Bridge architecture: a fresh EqV3 backbone with identical configuration is randomly initialized (seed 922806) and frozen, dual-parity blocks 2, 4, and 7 features are extracted under the identical right-handed inversion convention, and the identical 2,829,951-parameter bridge and student decoder are trained from scratch on the exact same 50,000 bulk crystals, probe sequence, and two-stage schedule. On the 1,000-crystal validation set, the random-encoder model stalls at $R_{\rm val}=6.3078\%$ versus $5.6368\%$ for pretrained EqV3. To verify seed-to-seed stability and rule out self-distillation artifacts, we additionally cross-check three paired random seeds (922806--922808) and EdenGNN-Uni self-distillation under matched data order and initialization.

\subsection{Sample selection and DFT reference calculations}

\paragraph{MoS$_2$ nanotubes and bulk crystals.} Four near-relaxed MoS$_2$ nanotubes ($n=17,28,36,39$) were held out throughout transfer training and checkpoint selection. Models are scored on the same 61,440 stratified sites per tube from archived VASP~\cite{pbVASP} PBE~\cite{pbPBE} PAW~\cite{pbPAW} charge grids (400~eV cutoff, $1\times1\times9$ $k$-mesh, non-spin-polarized, no SOC) using the same isolated-atom reference. Separate single-structure calibration fits 42 output-channel gains (21 per element, ridge penalty $10^{-3}$) on a fifth archived tube ($n=27$) with all neural weights frozen. For the bulk benchmark, we evaluate on all 965 tungsten-free crystals from a random 1,000-crystal draw of the 3,219 EdenGNN-Uni test structures with archived static DFT grids (35 tungsten structures required a \texttt{W\_pv} PAW reference unavailable locally), scored on native FFT grids without calibration.

\paragraph{One deformation per crystal (100 pairs).} We selected 100 distinct parent crystals from the Materials Project~\cite{pbMP} spanning eight elements and 42 chemical systems, disjoint from training, validation, and the ten-crystal scan below (two overlap with the 965-crystal bulk test). Twenty-five parents were assigned to each combination of tension/compression and uniaxial/biaxial strain, with strain amplitudes drawn uniformly from 2--10\% (realized 2.08--9.97\%) at fixed fractional coordinates. Strained densities were computed with VASP 6.3.2 matching each parent's PAW datasets, $k$-mesh, FFT grid dimensions, 520~eV cutoff, spin polarization, and tetrahedron integration with Bl{\"o}chl corrections~\cite{pbTet}, requiring integrated density agreement within $0.002$ electrons.

\paragraph{Matched strain scans (10 crystals, 200 strains).} To probe opposite strain signs and systematic amplitude dependence within each crystal, ten small-cell ($\le10$ atoms) tungsten-free parents were subjected to uniaxial and biaxial strains at $\pm1,\pm2,\pm3,\pm4,$ and $\pm5\%$ (200 deformed structures plus 10 matched zero-strain references), computed with static PAW--PBE, fixed fractional coordinates, no spin polarization, fixed $\Gamma$-centered $k$-meshes, 0.05~eV Gaussian smearing, $10^{-7}$~eV electronic convergence, and 520~eV cutoff (650~eV for LiTaO$_3$).

\paragraph{Full-periodic-table strain and atomic-rattle benchmark (579 crystals, 79 elements).} To evaluate transferability across the periodic table and test a distinct perturbation mode beyond homogeneous lattice strain, we drew held-out crystals from the unused EdenGNN-Uni test pool (disjoint from the 100-crystal strain test and spanning 1 to 42 atoms per cell, median 12) prioritized for maximum chemical-element diversity across all 79 elements with locally verified PBE PAW datasets. Each crystal was assigned to either homogeneous lattice strain (uniaxial or biaxial, tension or compression, realized magnitude $2.01\text{--}7.95\%$ at fixed fractional coordinates) or internal atomic rattling (fixed cell, zero-net-drift Gaussian Cartesian displacements with target root-mean-square amplitude $0.05\text{--}0.15$~\AA{} and minimum interatomic distance guarded above $\max(0.65\text{ \AA}, 0.78\,d_0)$). Both the unperturbed reference and perturbed geometry of every pair were computed consistently with static non-spin-polarized PAW--PBE in VASP 6.3.2 (\texttt{PREC=Accurate}, 520~eV plane-wave cutoff, $10^{-6}$~eV electronic convergence, $\Gamma$-centered $k$-mesh at $\approx 0.25$~\AA$^{-1}$ spacing, 0.05~eV Gaussian smearing, \texttt{LASPH=.TRUE.}), yielding 579 fully converged DFT pairs (290 lattice-strain pairs: 143 tension and 147 compression; 289 atomic-rattle pairs) across 79 distinct elements.

\subsection{Total-density and charge-redistribution errors}

Let $g$ index fractional grid sites in a cell of volume $V$ with normalized weights $w_g$ (uniform on full FFT grids, stratum weights on sampled nanotube grids). For model $m$, let $\rho^m_g$ and $\rho^{\rm DFT}_g$ denote predicted and reference total valence densities. The total-density error of one structure is
\begin{equation}
 E_{\rho}(m)=100\,\frac{V\sum_g w_g |\rho^m_g-\rho^{\rm DFT}_g|}{N_e},
\end{equation}
where $N_e$ is the reference valence electron count. For a perturbed cell labeled $u$, set $q^m_u(\mathbf s)=V_u\rho^m_u(\mathbf s)$ on fractional coordinate $\mathbf s$ and form $\Delta q^m_u=q^m_u-\mathcal I_u q^m_0$, where $0$ denotes the unperturbed crystal and $\mathcal I_u$ resamples its field onto the perturbed grid by Fourier interpolation. The charge-redistribution error normalized by the parent valence electron count $N_e$ is
\begin{equation}
 E_{\Delta,N_e}(m,u)=100\,\frac{\sum_g w_g|\Delta q^m_{u,g}-\Delta q^{\rm DFT}_{u,g}|}{N_e}.
\end{equation}
Both $E_\rho$ and $E_{\Delta,N_e}$ share the same valence-electron scale and give equal weight to each structure; at zero perturbation, $E_{\Delta,N_e}=0$ by construction. Checkpoint selection uses pooled deformation-density agreement with the frozen teacher on the 1,000 validation crystals,
\begin{equation}
 R_{\rm val}=100\sqrt{\frac{\sum_{i,g}(d^m_{ig}-d^{\rm T}_{ig})^2}{\sum_{i,g}(d^{\rm T}_{ig})^2}}.
\end{equation}

\section{Results and Discussion}

\subsection{Electron density in held-out \texorpdfstring{MoS$_2$}{MoS2} nanotubes}

\begin{figure}[!tb]
\centering
\includegraphics[width=8.3cm]{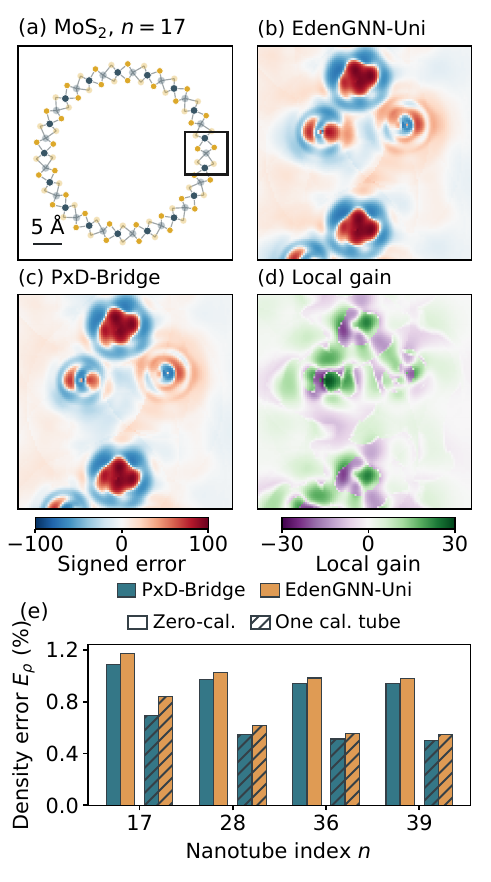}
\caption{Density errors on held-out MoS$_2$ nanotubes. (a) Axial view of the $n=17$ tube. Blue-gray and gold denote Mo and S; darker atoms lie in the midplane. The box marks the $8\times8$~\AA$^2$ window used in (b)--(d). (b,c) Zero-calibration signed errors $e_m=\rho_m-\rho_{\rm DFT}$ for EdenGNN-Uni and PxD-Bridge on the same grid (red: excess density; blue: deficit). (d) Local reduction in absolute error, $|e_{\rm teacher}|-|e_{\rm student}|$ (green favors PxD-Bridge; purple favors EdenGNN-Uni). Maps use symmetric $\mathrm{asinh}$ scales in $10^{-3}$~electrons/\AA$^3$ with a shared scale for (b,c). (e) Total-density errors for PxD-Bridge (teal) and EdenGNN-Uni (orange) sharing a common zero baseline: solid bars show zero-calibration predictions and hatched bars show predictions after calibrating 42 output gains per model on the same $n=27$ tube.}\label{fig:nanotube-transfer}
\end{figure}

\paragraph{Zero-calibration transfer.} Trained solely on bulk teacher densities, PxD-Bridge predicts the total valence density more accurately than EdenGNN-Uni on all four held-out MoS$_2$ nanotubes without calibration [Fig.~\ref{fig:nanotube-transfer}(e) and Table~\ref{tab:tubevariants}]. The four-tube mean error drops from $1.0408\%$ for EdenGNN-Uni to $0.9849\%$ for PxD-Bridge, improving on EdenGNN-Uni at every diameter ($1.0874\%$ versus $1.1740\%$ at $n=17$; $0.9722\%$ versus $1.0263\%$ at $n=28$; $0.9404\%$ versus $0.9826\%$ at $n=36$; $0.9397\%$ versus $0.9802\%$ at $n=39$). Transverse midplane error maps at the highest curvature ($n=17$) show that both models concentrate residuals around the curved Mo--S wall [Fig.~\ref{fig:nanotube-transfer}(a)--(c)], while the local gain map confirms net error reduction across the tube cross section [Fig.~\ref{fig:nanotube-transfer}(d)].

\paragraph{Single-structure calibration and parity ablation.} Fitting 42 output-channel gains per model on the fifth archived tube ($n=27$) reduces the four-tube mean error from $0.6383\%$ for EdenGNN-Uni to $0.5645\%$ for PxD-Bridge (Table~\ref{tab:tubevariants}). PxD-Bridge improves on the calibrated teacher at every held-out diameter ($0.6937\%$ versus $0.8399\%$ at $n=17$, a 0.1462-point gain; $0.5495\%$ versus $0.6133\%$ at $n=28$; $0.5117\%$ versus $0.5560\%$ at $n=36$; and $0.5030\%$ versus $0.5441\%$ at $n=39$, a 0.0411-point gain). Removing the unnatural-parity EqV3 channel (without $X^-$) raises the mean error on all four tubes both before calibration ($1.0083\%$ versus $0.9849\%$) and after calibration ($0.5681\%$ versus $0.5645\%$), confirming that the dual-parity interface aids out-of-distribution generalization to curved 1D walls.

\begin{table}[tbp]\centering
\caption{Mean MoS$_2$ nanotube total-density error $E_\rho$ (\%) against DFT across the four held-out diameters ($n=17,28,36,39$) before calibration (Zero) and after fitting 42 channel gains on $n=27$ (One).}\label{tab:tubevariants}
\begin{ruledtabular}\begin{tabular}{lrr}
Model & Zero & One ($n=27$) \\ \hline
EdenGNN-Uni (teacher) & 1.0408 & 0.6383 \\
PxD-Bridge (without $X^-$) & 1.0083 & 0.5681 \\
\textbf{PxD-Bridge} & \textbf{0.9849} & \textbf{0.5645} \\
\end{tabular}\end{ruledtabular}\end{table}

\subsection{Bulk electron density and its response to strain}

\paragraph{Total density on 965 held-out bulk crystals.} Evaluated on the 965 tungsten-free bulk test crystals on native DFT grids without calibration, PxD-Bridge lowers the mean total-density error relative to its teacher, from $1.2463\%$ for EdenGNN-Uni to $1.2407\%$ ($\Delta=-0.0055$ percentage points, 95\% paired bootstrap interval $-0.0114$ to $+0.0001$; lower on 439/965 crystals). Thus, a student trained strictly on EdenGNN-Uni's bulk teacher targets matches and slightly surpasses its teacher even on held-out static bulk densities.

\begin{figure}[tbp]
\centering
\includegraphics[width=8.3cm]{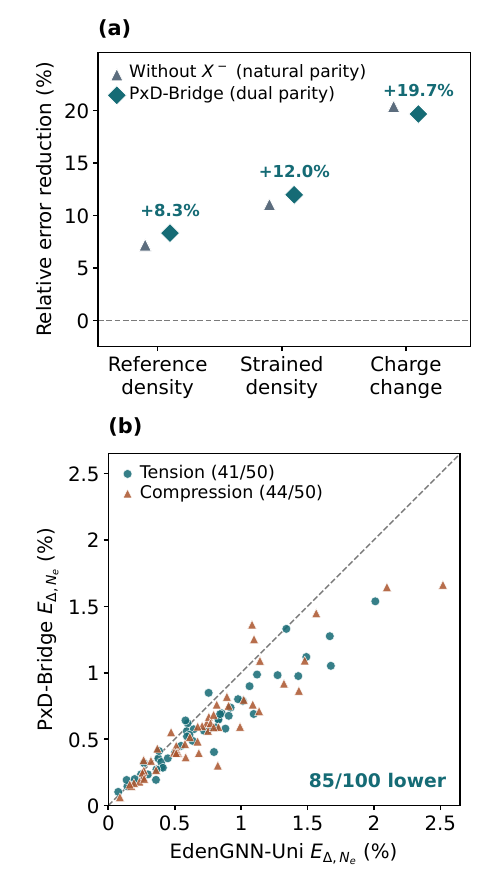}
\caption{Zero-calibration density and charge-redistribution errors on 100 crystals with one random deformation each (2--10\% strain). (a) Relative error reduction, $100(1-\bar E_{\rm student}/\bar E_{\rm teacher})$, for the natural-parity ablation (without $X^-$) and the full dual-parity PxD-Bridge model across unstrained reference density, strained density, and charge redistribution. Positive values favor the student: PxD-Bridge delivers $+8.3\%$ on reference density, $+12.0\%$ on strained density, and $+19.7\%$ on charge redistribution. (b) Crystal-by-crystal comparison of $E_{\Delta,N_e}$ (\% of valence electrons) for EdenGNN-Uni versus PxD-Bridge. Circles mark tension and triangles compression; PxD-Bridge lies below the equality line on 85 of the 100 crystals.}\label{fig:strain100-paired}
\end{figure}

\paragraph{One deformation per parent crystal (100 pairs).} Table~\ref{tab:locked-summary} and Fig.~\ref{fig:strain100-paired} summarize zero-calibration performance on the 100-crystal strain test, where the integrated absolute DFT charge redistribution averages $5.312\%$ of the valence electron count. PxD-Bridge improves decisively over EdenGNN-Uni across all three metrics [Fig.~\ref{fig:strain100-paired}(a)]:
\begin{itemize}
 \item \emph{Unstrained total density $E_\rho(0)$}: drops from $1.143\%$ to $1.048\%$ ($-0.095$ percentage points, 95\% paired interval $-0.158$ to $-0.047$; lower on 70/100 crystals).
 \item \emph{Strained total density $E_\rho(u)$}: drops from $1.402\%$ to $1.234\%$ ($-0.168$ percentage points, 95\% interval $-0.230$ to $-0.118$; lower on 80/100 crystals).
 \item \emph{Strain-induced redistribution $E_{\Delta,N_e}$}: drops from $0.739\%$ to $0.594\%$---a $19.7\%$ relative error reduction ($-0.145$ percentage points, 95\% interval $-0.180$ to $-0.112$; lower on 85/100 crystals) [Fig.~\ref{fig:strain100-paired}(b)].
\end{itemize}
Excluding the two crystals shared with the 965-crystal bulk benchmark gives virtually identical redistribution performance (gain $0.148$ percentage points, 84 wins among 98 crystals). Importantly, PxD-Bridge delivers strongly significant gains of nearly equal magnitude under both deformation signs: under tension (50 pairs), $E_{\Delta,N_e}$ falls from $0.726\%$ to $0.583\%$ (gain $0.143$ points, 95\% interval $0.102$ to $0.187$, 41/50 wins); under compression (50 pairs), it falls from $0.752\%$ to $0.604\%$ (gain $0.148$ points, 95\% interval $0.097$ to $0.204$, 44/50 wins). Comparing against the matched natural-parity ablation (without $X^-$) shows that adding $X_i^-$ lowers both unstrained ($1.048\%$ versus $1.061\%$; 70 versus 62 wins over EdenGNN-Uni) and strained ($1.234\%$ versus $1.247\%$; 80 versus 76 wins) total-density errors while maintaining the same 85/100 redistribution win rate ($0.594\%$ versus $0.588\%$).

\begin{table*}[tbp]
\centering
\caption{Zero-calibration benchmark summary (\% of total valence electron count) comparing the EdenGNN-Uni teacher, the matched natural-parity ablation (without $X^-$), and the full dual-parity PxD-Bridge model across all locked DFT benchmarks plus official 1,000-crystal teacher validation ($R_{\rm val}$). Win counts report crystals where PxD-Bridge achieves strictly lower error than EdenGNN-Uni.}\label{tab:locked-summary}
\begin{ruledtabular}
\begin{tabular}{lrrrr}
Benchmark & EdenGNN-Uni & Without $X^-$ & \textbf{PxD-Bridge} & Wins vs.\ Eden \\ \hline
Teacher validation $R_{\rm val}$ (1,000 crystals) & -- & 5.7978 & \textbf{5.6368} & -- \\
Bulk total density $E_\rho$ (965 non-W crystals) & 1.2463 & -- & \textbf{1.2407} & 439/965 \\
Unstrained total density $E_\rho(0)$ (100 crystals) & 1.1433 & 1.0612 & \textbf{1.0483} & 70/100 \\
Strained total density $E_\rho(u)$ (100 crystals) & 1.4017 & 1.2471 & \textbf{1.2341} & 80/100 \\
Strain redistribution $E_{\Delta,N_e}$ (100 pairs) & 0.7390 & 0.5884 & \textbf{0.5937} & 85/100 \\
\quad Tension subset (50 pairs) & 0.7259 & 0.5719 & \textbf{0.5832} & 41/50 \\
\quad Compression subset (50 pairs) & 0.7521 & 0.6048 & \textbf{0.6042} & 44/50 \\
MoS$_2$ nanotubes zero-cal $E_\rho$ (4 tubes) & 1.0408 & 1.0083 & \textbf{0.9849} & 4/4 \\
\end{tabular}
\end{ruledtabular}
\end{table*}

\begin{figure}[!htbp]
\centering\includegraphics[width=8.3cm]{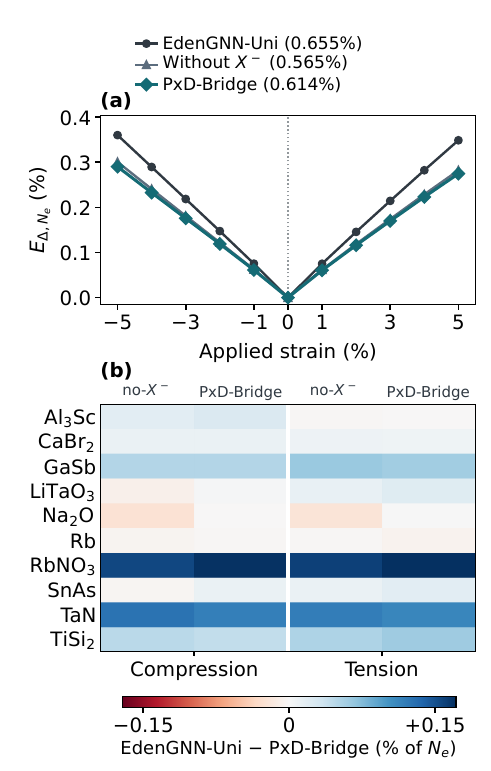}
\caption{Charge redistribution along matched strain paths in ten crystals. (a) At each nonzero signed strain ($\pm1\%$ to $\pm5\%$), points give mean $E_{\Delta,N_e}$ over ten crystals and two strain modes (uniaxial and biaxial) relative to each crystal's zero-strain reference. Legend values in parentheses give each model's mean zero-strain total-density error $E_\rho(0)$ across the ten unstrained references. (b) For each material and strain sign, color gives the mean reduction in $E_{\Delta,N_e}$ relative to EdenGNN-Uni averaged over five amplitudes and both modes (blue favors PxD-Bridge; red favors EdenGNN-Uni). The two columns within each sign compare the natural-parity ablation (no-$X^-$) and the full dual-parity PxD-Bridge model.}\label{fig:strain200}
\end{figure}

\begin{table}[!tbp]
\centering
\caption{Zero-calibration errors on the full 200-strain grid (10 crystals $\times$ 5 amplitudes $\times$ 2 modes per sign). Each entry averages 100 strained structures at one strain sign, expressed as a percentage of the total valence electron count $N_e$. The final column counts matched compression--tension pairs (out of 100) where the student lowers $E_{\Delta,N_e}$ relative to EdenGNN-Uni for both signs.}\label{tab:strain200}
\begin{ruledtabular}\begin{tabular}{lrrrrr}
& \multicolumn{2}{c}{$E_\rho$} & \multicolumn{2}{c}{$E_{\Delta,N_e}$} & Both \\
Model & Compr. & Tens. & Compr. & Tens. & signs \\ \hline
EdenGNN-Uni (teacher) & 0.697 & 0.704 & 0.2182 & 0.2132 & -- \\
PxD-Bridge (without $X^-$) & 0.585 & 0.597 & 0.1814 & 0.1731 & 59/100 \\
\textbf{PxD-Bridge} & \textbf{0.641} & \textbf{0.634} & \textbf{0.1758} & \textbf{0.1689} & \textbf{76/100} \\
\end{tabular}\end{ruledtabular}
\end{table}

\paragraph{Compression and tension at matched amplitudes (10 crystals, 200 strains).} The ten-crystal scan evaluates opposite deformations of the same material at equal strain magnitudes ($\pm1\%$ to $\pm5\%$, uniaxial and biaxial) without calibration. Along each path, the energy slope at zero strain probes the stress projection, while the same bond-length and angle changes alter orbital overlap and redistribute valence charge. Figure~\ref{fig:strain200} and Table~\ref{tab:strain200} compare EdenGNN-Uni, the natural-parity ablation (without $X^-$), and PxD-Bridge across all 100 matched compression--tension pairs.

PxD-Bridge achieves the lowest mean redistribution error at every nonzero signed strain amplitude [Fig.~\ref{fig:strain200}(a)], reducing mean $E_{\Delta,N_e}$ from $0.2182\%$ to $0.1758\%$ under compression and from $0.2132\%$ to $0.1689\%$ under tension, while also lowering mean total-density error $E_\rho$ under both compression ($0.641\%$ versus $0.697\%$) and tension ($0.634\%$ versus $0.704\%$) (Table~\ref{tab:strain200}). Resampling the ten parent crystals together (20,000 bootstrap draws), PxD-Bridge yields 95\% crystal-level confidence intervals that strictly exclude zero for \emph{both} deformation signs: $+0.0424$ percentage points under compression (95\% interval $+0.0124$ to $+0.0794$, lower on 76/100 structures and 8 of 10 crystals) and $+0.0443$ points under tension (95\% interval $+0.0137$ to $+0.0818$, lower on 79/100 structures and 8 of 10 crystals). Moreover, retaining the unnatural-parity EqV3 channel $X_i^-$ proves decisive for opposite-sign consistency: PxD-Bridge outperforms EdenGNN-Uni for both compression and tension simultaneously on 76 of the 100 matched pairs, compared with 59/100 when $X_i^-$ is omitted, removing the teacher's residual advantage on LiTaO$_3$, CaBr$_2$, Al$_3$Sc (compression), and Na$_2$O (tension) [Fig.~\ref{fig:strain200}(b)].

\subsection{Full-periodic-table strain and atomic-rattle benchmark (579 crystals)}
\label{sec:bench500}

\begin{figure*}[tbp]
\centering
\includegraphics[width=17.78cm]{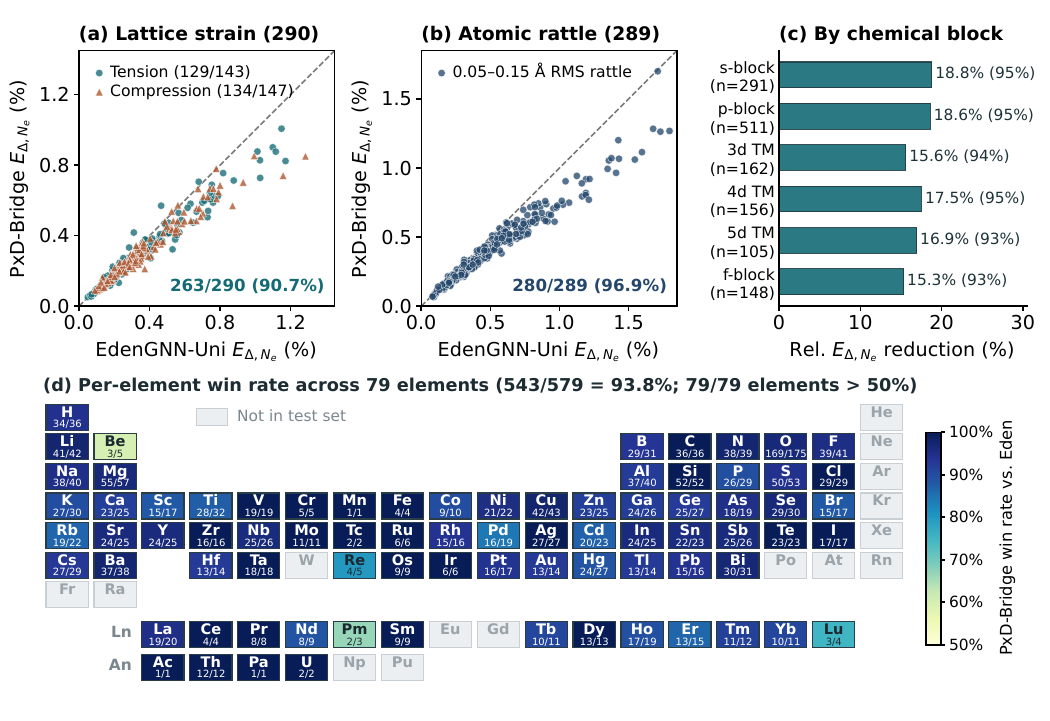}
\caption{Zero-calibration transfer across 579 held-out crystals spanning 79 elements of the periodic table. (a) Crystal-by-crystal comparison of charge-redistribution error $E_{\Delta,N_e}$ (\% of valence electrons) for EdenGNN-Uni versus PxD-Bridge on 290 lattice-strain pairs (2--8\% uniaxial or biaxial strain; tension circles and compression triangles): PxD-Bridge is more accurate on 263/290 crystals ($90.7\%$). (b) Corresponding comparison on 289 fixed-cell atomic-rattle pairs ($0.05\text{--}0.15$~\AA{} RMS displacements): PxD-Bridge is more accurate on 280/289 crystals ($96.9\%$). (c) Relative reduction in mean $E_{\Delta,N_e}$ versus EdenGNN-Uni across all six chemical blocks ($s$-block, $p$-block, $3d/4d/5d$ transition metals, and $f$-block), with per-block win rates in parentheses ($93\text{--}95\%$). (d) Periodic-table heatmap of per-element win rates on charge redistribution (tile fractions show PxD-Bridge wins over crystals containing that element): PxD-Bridge achieves $>50\%$ win rate on all 79/79 tested elements and $\ge 80\%$ on 76/79 elements.}\label{fig:bench500}
\end{figure*}

The 100-crystal strain benchmark spans eight elements and tests only homogeneous lattice deformations. To test whether the transfer advantage generalizes across the periodic table and extends to internal atomic displacements (a perturbation directly probed by interatomic-potential forces during OMat24 pretraining), we evaluated EdenGNN-Uni, the natural-parity ablation (without $X^-$), and PxD-Bridge without calibration on all 579 converged DFT pairs spanning 79 distinct elements (Table~\ref{tab:bench500} and Fig.~\ref{fig:bench500}).

Across all 579 crystals, PxD-Bridge outperforms EdenGNN-Uni on every observable (Table~\ref{tab:bench500}). On unperturbed static total density $E_\rho(0)$, PxD-Bridge lowers the mean error from $0.6852\%$ to $0.6760\%$ ($\Delta=-0.0092$ percentage points, 95\% paired bootstrap interval $-0.0173$ to $-0.0016$). After perturbation, the advantage widens sharply: mean perturbed total-density error $E_\rho(u)$ drops from $0.8367\%$ to $0.7736\%$ ($\Delta=-0.0631$ percentage points, 95\% interval $-0.0743$ to $-0.0528$; lower on 433/579 crystals, $74.8\%$). For the perturbation-induced charge redistribution $E_{\Delta,N_e}$, PxD-Bridge lowers the mean error by $17.7\%$, from $0.4669\%$ to $0.3841\%$ ($\Delta=-0.0828$ percentage points, 95\% interval $-0.0909$ to $-0.0750$), outperforming EdenGNN-Uni on 543 of the 579 crystals ($93.8\%$).

Separating the two perturbation arms reveals an especially striking result for internal atomic displacements [Fig.~\ref{fig:bench500}(a,b)]:
\begin{itemize}
 \item \emph{Lattice strain (290 crystals; mean DFT redistribution $5.26\%$ of $N_e$)}: PxD-Bridge lowers $E_{\Delta,N_e}$ by $15.2\%$, from $0.3938\%$ to $0.3338\%$, winning on 263/290 crystals ($90.7\%$; 95\% interval $-0.0688$ to $-0.0519$ percentage points), with symmetric accuracy under tension (129/143, $90.2\%$) and compression (134/147, $91.2\%$).
 \item \emph{Atomic rattle (289 crystals; mean DFT redistribution $14.89\%$ of $N_e$)}: internal displacements generate nearly triple the charge redistribution of homogeneous strain, and here the force-pretrained encoder shines brightest. PxD-Bridge cuts perturbed total-density error from $0.8619\%$ to $0.7737\%$ (247/289 wins, $85.5\%$; 95\% interval $-0.1039$ to $-0.0738$) and reduces redistribution error $E_{\Delta,N_e}$ by $19.6\%$, from $0.5402\%$ to $0.4345\%$ (95\% interval $-0.1187$ to $-0.0934$), surpassing EdenGNN-Uni on 280 of 289 crystals ($96.9\%$).
\end{itemize}

The transfer gain is uniform across the entire periodic table rather than concentrated in a few chemistries [Fig.~\ref{fig:bench500}(c,d)]. Grouping crystals by chemical block yields $15.3\text{--}18.8\%$ relative redistribution error reductions and $93.2\text{--}95.3\%$ win rates in every block: $s$-block ($18.8\%$, 275/291), $p$-block ($18.6\%$, 487/511), $3d$ transition metals ($15.6\%$, 153/162), $4d$ ($17.5\%$, 148/156), $5d$ ($16.9\%$, 98/105), and $f$-block lanthanides and actinides ($15.3\%$, 138/148). At the individual-element level, PxD-Bridge wins on $>50\%$ of crystals for all 79/79 tested elements, $\ge 80\%$ for 76/79 elements, and $100\%$ for 25 elements (including C, Si, Cl, V, Cr, Fe, Mo, Ru, Ag, Te, Ta, Os, Ir, and Th). On this 579-crystal bulk panel, the dual-parity and natural-parity models perform comparably on average ($0.3841\%$ versus $0.3839\%$), whereas the unnatural-parity channel $X_i^-$ yields its clearest incremental gains on teacher validation ($R_{\rm val}$), curved MoS$_2$ nanotubes, and opposite-sign consistency in the matched strain scan.

\begin{table*}[tbp]
\centering
\caption{Zero-calibration errors (\% of total valence electron count $N_e$) on the 579-crystal full-periodic-table perturbation benchmark spanning 79 elements (290 lattice-strain pairs + 289 atomic-rattle pairs). Win counts report crystals where PxD-Bridge achieves strictly lower error than EdenGNN-Uni.}\label{tab:bench500}
\begin{ruledtabular}
\begin{tabular}{llrrrr}
Subset & Metric & EdenGNN-Uni & Without $X^-$ & \textbf{PxD-Bridge} & Wins vs.\ Eden \\ \hline
All ($N=579$, 79 els.) & Unperturbed $E_\rho(0)$ & 0.6852 & 0.6762 & \textbf{0.6760} & 288/579 \\
 & Perturbed $E_\rho(u)$ & 0.8367 & 0.7735 & \textbf{0.7736} & 433/579 (74.8\%) \\
 & Redistribution $E_{\Delta,N_e}$ & 0.4669 & 0.3839 & \textbf{0.3841} & 543/579 (93.8\%) \\ \hline
Lattice strain ($N=290$) & Unperturbed $E_\rho(0)$ & 0.7111 & 0.7036 & \textbf{0.7023} & 152/290 \\
 & Perturbed $E_\rho(u)$ & 0.8115 & 0.7743 & \textbf{0.7734} & 186/290 (64.1\%) \\
 & Redistribution $E_{\Delta,N_e}$ & 0.3938 & 0.3332 & \textbf{0.3338} & 263/290 (90.7\%) \\
 & \quad Tension ($N=143$) & 0.3916 & 0.3340 & \textbf{0.3349} & 129/143 (90.2\%) \\
 & \quad Compression ($N=147$) & 0.3960 & 0.3324 & \textbf{0.3328} & 134/147 (91.2\%) \\ \hline
Atomic rattle ($N=289$) & Unperturbed $E_\rho(0)$ & 0.6592 & 0.6486 & \textbf{0.6496} & 136/289 \\
 & Perturbed $E_\rho(u)$ & 0.8619 & 0.7727 & \textbf{0.7737} & 247/289 (85.5\%) \\
 & Redistribution $E_{\Delta,N_e}$ & 0.5402 & 0.4348 & \textbf{0.4345} & 280/289 (96.9\%) \\
\end{tabular}
\end{ruledtabular}
\end{table*}

\subsection{Contribution of interatomic-potential pretraining}

\begin{figure}[!tbp]
\centering
\includegraphics[width=8.3cm]{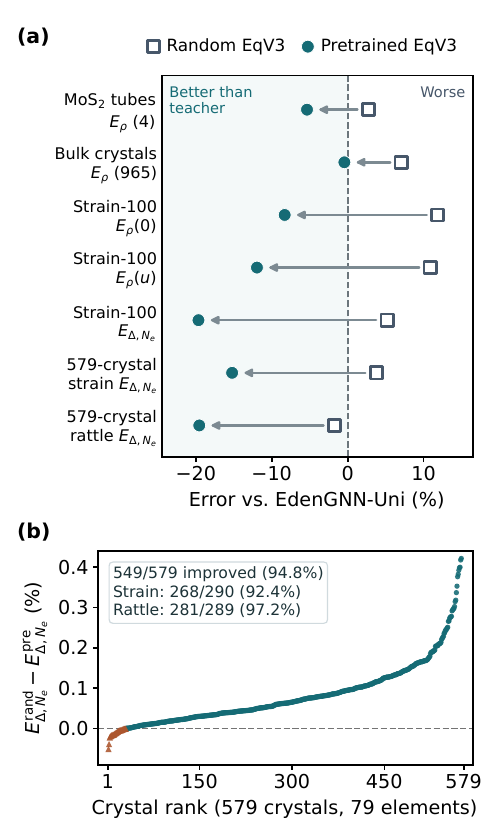}
\caption{Matched dual-parity EqV3 pretraining ablation across all zero-calibration benchmarks. (a) Relative error difference versus EdenGNN-Uni, $100(\bar E_{\rm student}/\bar E_{\rm teacher}-1)$, for identical dual-parity PxD-Bridge models driven by frozen randomly initialized EqV3 (open squares) versus pretrained EqV3 OMat-G (filled teal circles). Negative values (shaded left) favor the student over the teacher. Without energy-and-force pretraining, the identical bridge trails EdenGNN-Uni across static and strain benchmarks; enabling OMat24 pretraining drives every arrow across the teacher baseline. (b) Ranked per-crystal redistribution gain ($E_{\Delta,N_e}^{\rm rand}-E_{\Delta,N_e}^{\rm pre}$) across all 579 full-periodic-table crystals (79 elements): pretraining lowers the error on 549/579 crystals ($94.8\%$; 268/290 strain, 281/289 atomic rattle).}\label{fig:pretraining-pairs}
\end{figure}

How much of the accuracy gain arises specifically from energy-and-force pretraining in EqV3 OMat-G, rather than from the dual-parity bridge architecture, short-range bond moments, or teacher distillation alone? Table~\ref{tab:pretraining-dual} and Fig.~\ref{fig:pretraining-pairs} answer this directly by comparing frozen pretrained EqV3 OMat-G against a frozen randomly initialized EqV3 backbone inside the \emph{exact same} 2,829,951-parameter dual-parity PxD-Bridge architecture, trained on the exact same 50,000 bulk crystals, seed (922806), and two-stage schedule.

Without energy-and-force pretraining, the identical dual-parity bridge architecture \emph{fails to surpass EdenGNN-Uni} across static densities and strain redistribution (Table~\ref{tab:pretraining-dual} and Fig.~\ref{fig:pretraining-pairs}(a)). The random-EqV3 bridge lags the teacher on all four MoS$_2$ nanotubes ($1.0692\%$ versus $1.0408\%$, 0/4 wins), on the 965 bulk crystals ($1.3341\%$ versus $1.2463\%$, only 83/965 wins), on the 100-crystal strain benchmark at zero strain ($1.2788\%$ versus $1.1433\%$), under strain ($1.5540\%$ versus $1.4017\%$), and on strain redistribution ($0.7773\%$ versus $0.7390\%$, $+0.0383$ percentage points worse, 95\% interval $+0.0054$ to $+0.0726$), as well as across the 579-crystal full-periodic-table benchmark on unperturbed density ($0.7590\%$ versus $0.6852\%$), perturbed density ($0.8945\%$ versus $0.8367\%$), and overall redistribution ($0.4696\%$ versus $0.4669\%$). Even on internal atomic rattling, where the short-range bond moments and neighbor update bring the random-EqV3 model close to the teacher ($0.5308\%$ versus $0.5402\%$), it still falls far short of pretrained PxD-Bridge ($0.4345\%$).

Enabling OMat24 energy-and-force pretraining transforms performance uniformly across every benchmark and observable [Fig.~\ref{fig:pretraining-pairs}(a)]:
\begin{itemize}
 \item \emph{Static bulk and nanotube densities}: pretrained EqV3 beats random EqV3 on all 4/4 MoS$_2$ nanotubes ($0.9849\%$ versus $1.0692\%$), on 877/965 bulk crystals ($90.9\%$; $\Delta=-0.0934$ percentage points, 95\% interval $-0.1004$ to $-0.0868$), on 87/100 unstrained parents ($-0.2305$ points, interval $-0.2981$ to $-0.1765$), and on 515/579 unperturbed full-periodic-table crystals ($88.9\%$; $-0.0830$ points, interval $-0.0924$ to $-0.0742$).
 \item \emph{Perturbed total densities}: pretrained EqV3 outperforms random EqV3 on 91/100 strained crystals ($1.2341\%$ versus $1.5540\%$; $-0.3199$ points, interval $-0.3929$ to $-0.2598$) and on 533/579 perturbed full-periodic-table crystals ($92.1\%$; $-0.1209$ points, interval $-0.1336$ to $-0.1095$).
 \item \emph{Charge redistribution}: pretrained EqV3 improves over random EqV3 on 95/100 crystals in the 100-pair strain benchmark ($0.5937\%$ versus $0.7773\%$; $-0.1836$ points, interval $-0.2174$ to $-0.1517$) and on 549/579 crystals ($94.8\%$) across the 79-element benchmark [Fig.~\ref{fig:pretraining-pairs}(b)]---specifically 268/290 ($92.4\%$; $-0.0748$ points, interval $-0.0835$ to $-0.0665$) under lattice strain and 281/289 ($97.2\%$; $-0.0963$ points, interval $-0.1104$ to $-0.0835$) under atomic rattling.
\end{itemize}

\begin{table*}[tbp]
\centering
\caption{Matched dual-parity EqV3 pretraining ablation across all zero-calibration benchmarks (\% of total valence electron count $N_e$) plus official 1,000-crystal teacher validation ($R_{\rm val}$). Random and pretrained EqV3 backbones use the identical 2,829,951-parameter dual-parity bridge, 50,000 bulk teacher structures, seed (922806), and two-stage training schedule. The final column gives paired win counts where pretrained EqV3 achieves strictly lower error than random EqV3.}\label{tab:pretraining-dual}
\begin{ruledtabular}
\begin{tabular}{lrrrr}
Benchmark & EdenGNN-Uni & Random EqV3 & \textbf{Pretrained EqV3} & Pre.\ vs.\ Rand. \\ \hline
Teacher validation $R_{\rm val}$ (1,000 crystals) & -- & 6.3078 & \textbf{5.6368} & -- \\
MoS$_2$ nanotubes zero-cal $E_\rho$ (4 tubes) & 1.0408 & 1.0692 & \textbf{0.9849} & 4/4 (100\%) \\
Bulk total density $E_\rho$ (965 non-W crystals) & 1.2463 & 1.3341 & \textbf{1.2407} & 877/965 (90.9\%) \\ \hline
Strain-100 unstrained $E_\rho(0)$ (100 crystals) & 1.1433 & 1.2788 & \textbf{1.0483} & 87/100 (87.0\%) \\
Strain-100 strained $E_\rho(u)$ (100 crystals) & 1.4017 & 1.5540 & \textbf{1.2341} & 91/100 (91.0\%) \\
Strain-100 redistribution $E_{\Delta,N_e}$ (100 pairs) & 0.7390 & 0.7773 & \textbf{0.5937} & 95/100 (95.0\%) \\ \hline
579-crystal unperturbed $E_\rho(0)$ (579 crystals) & 0.6852 & 0.7590 & \textbf{0.6760} & 515/579 (88.9\%) \\
579-crystal perturbed $E_\rho(u)$ (579 crystals) & 0.8367 & 0.8945 & \textbf{0.7736} & 533/579 (92.1\%) \\
579-crystal redistribution $E_{\Delta,N_e}$ (579 pairs) & 0.4669 & 0.4696 & \textbf{0.3841} & 549/579 (94.8\%) \\
\quad Lattice strain subset (290 pairs) & 0.3938 & 0.4086 & \textbf{0.3338} & 268/290 (92.4\%) \\
\quad Atomic rattle subset (289 pairs) & 0.5402 & 0.5308 & \textbf{0.4345} & 281/289 (97.2\%) \\
\end{tabular}
\end{ruledtabular}
\end{table*}

\paragraph{Seed-to-seed stability and self-distillation check.} Auxiliary three-seed controls (seeds 922806--922808) confirm that this pretraining advantage is insensitive to random initialization and does not arise from teacher self-distillation. On the 50-configuration strain subset, distilling EdenGNN-Uni's own frozen hidden states across three seeds yields a mean redistribution error of $0.2563\%$, remaining virtually identical to the teacher ($0.2545\%$). Likewise, pairing frozen random and pretrained EqV3 encoders across all three seeds on the 100-crystal strain panel reproduces a tight, consistent pretraining gain ($0.713\pm0.032\%$ versus $0.902\pm0.015\%$, with pretrained EqV3 lower on 91/100 crystals after averaging over seeds), matching the single-seed dual-parity gap ($0.594\%$ versus $0.777\%$, 95/100 wins) in both direction and magnitude.

\section{Conclusion}

We have tested whether a universal interatomic potential pretrained on abundant energies and forces can serve as a representation donor for a universal electron-density model. The transfer setting was deliberately strict: EdenGNN-Uni supplied all hidden-state, expansion-coefficient, and density supervision on its original 50,000 bulk structures, checkpoint selection used only teacher agreement, and no new DFT density labels entered training.

By connecting frozen EqV3 OMat-G to the EdenGNN-Uni decoder through a dual-parity $O(3)$-equivariant bridge---combining per-degree normalized blocks 2, 4, and 7, zero-extra-cost unnatural-parity ($X^-$) direct skips, short-range learned bond moments, multi-path Clebsch--Gordan couplings across all 13 active irreps, and a non-saturating Bessel/$\mathrm{asinh}$ neighbor update---PxD-Bridge surpasses its EdenGNN-Uni teacher across every zero-calibration DFT benchmark. Mean total-density errors decrease on 965 held-out bulk crystals ($1.241\%$ versus $1.246\%$), on all four held-out MoS$_2$ nanotubes ($0.985\%$ versus $1.041\%$), and on 100 separate crystals before ($1.048\%$ versus $1.143\%$) and after ($1.234\%$ versus $1.402\%$) lattice strain. On strain-induced charge redistribution, PxD-Bridge lowers mean $E_{\Delta,N_e}$ by $19.7\%$, from $0.739\%$ to $0.594\%$ of the valence electron count, winning on 85 of 100 crystals with statistically significant improvements under both tension and compression. Across an expanded 579-crystal full-periodic-table benchmark spanning 79 elements, this advantage extends robustly to both homogeneous strain (263/290 wins, $90.7\%$) and fixed-cell atomic rattling (280/289 wins, $96.9\%$), delivering a $17.7\%$ overall reduction in redistribution error ($0.384\%$ versus $0.467\%$, 543/579 wins) and a $>50\%$ win rate across all 79 tested elements. A matched control replacing pretrained EqV3 with a frozen randomly initialized EqV3 encoder inside the identical dual-parity bridge architecture fails to surpass EdenGNN-Uni, whereas pretrained EqV3 improves over the random encoder on 877/965 bulk crystals ($90.9\%$), 95/100 strained pairs, and 549/579 full-periodic-table crystals ($94.8\%$), confirming that energy-and-force pretraining drives the transfer gain.

These findings establish a practical route for scarce electronic-field observables: universal interatomic potentials trained on massive energy-and-force datasets can act not only as force engines, but as rich equivariant representation sources that elevate downstream density models beyond the teacher from which they were distilled.

\section*{Code and model availability}
The installable Python package, inference CLI, benchmark tables, and SHA-256-verified weights for PxD-Bridge (EqV3 OMat-G--EdenGNN-Uni) are publicly available at \href{https://github.com/klsk007/PxD-Bridge/releases/tag/v2.0.0}{\nolinkurl{github.com/klsk007/PxD-Bridge}} (release~\texttt{v2.0.0}). The release packages the trained 64-channel dual-parity bridge and student decoder (2,829,951 trainable parameters; checkpoint SHA-256~\texttt{10a4f286...1aeb7f45}) together with an upstream EqV3 OMat-G loading pipeline and a verified Si reference prediction (maximum absolute difference $5.2\times10^{-8}$~electrons/\AA$^3$ against the research implementation over all 1,728 grid points).

\FloatBarrier
\begin{acknowledgments}
We would like to thank Shandong Institute of Advanced Technology and National Supercomputing Center (Shuguang) for providing computational resources. This work is supported by the National Natural Science Foundation of Shandong Province (Grant No.~ZR2024QA040) and the National Natural Science Foundation of China (Grant No.~12604277). Xin Chen thanks the China Scholarship Council for financial support (No.~201606220031). Duo Wang acknowledges financial support from the Science and Technology Development Fund of Macao SAR (Nos.~0062/2023/ITP2 and 0016/2025/RIA1) and Macao Polytechnic University (Grant No.~RP/FCA-03/2023). AI assistance was used for language polishing and scripting; all AI-assisted text and code were reviewed and tested by the authors, who take full responsibility for the final manuscript.
\end{acknowledgments}

\renewcommand{\bibnumfmt}[1]{\textcolor{blue}{[#1]}}
\bibliography{physbridge_refs,potential_families}

@PREAMBLE{
 "\providecommand{\noopsort}[1]{}" 
 # "\providecommand{\singleletter}[1]{#1}%" 
}

@article{pbM3G,
  author = {Chen, Chi and Ong, Shyue Ping},
  title = {{A universal graph deep learning interatomic potential for the periodic table}},
  journal = {Nat Comput Sci},
  year = {2022},
  volume = {2},
  pages = {718--728},
  doi = {10.1038/s43588-022-00349-3}
}

@article{pbCHG,
  author = {Deng, Bowen and Zhong, Peichen and Jun, KyuJung and Riebesell, Janosh and Han, Kevin and Bartel, Christopher J. and Ceder, Gerbrand},
  title = {{CHGNet as a pretrained universal neural network potential for charge-informed atomistic modelling}},
  journal = {Nat Mach Intell},
  year = {2023},
  volume = {5},
  pages = {1031--1041},
  doi = {10.1038/s42256-023-00716-3}
}

@article{pbSeven,
  author = {Park, Yutack and Kim, Jaesun and Hwang, Seungwoo and Han, Seungwu},
  title = {{Scalable Parallel Algorithm for Graph Neural Network Interatomic Potentials in Molecular Dynamics Simulations}},
  journal = {J. Chem. Theory Comput.},
  year = {2024},
  volume = {20},
  pages = {4857--4868},
  doi = {10.1021/acs.jctc.4c00190}
}

@article{pbDPA,
  author = {Zhang, Duo and Liu, Xinzijian and Zhang, Xiangyu and Zhang, Chengqian and Cai, Chun and Bi, Hangrui and Du, Yiming and Qin, Xuejian and Peng, Anyang and Huang, Jiameng and Li, Bowen and Shan, Yifan and Zeng, Jinzhe and Zhang, Yuzhi and Liu, Siyuan and Li, Yifan and Chang, Junhan and Wang, Xinyan and Zhou, Shuo and Liu, Jianchuan and Luo, Xiaoshan and Wang, Zhenyu and Jiang, Wanrun and Wu, Jing and Yang, Yudi and Yang, Jiyuan and Yang, Manyi and Gong, Fu-Qiang and Zhang, Linshuang and Shi, Mengchao and Dai, Fu-Zhi and York, Darrin M. and Liu, Shi and Zhu, Tong and Zhong, Zhicheng and Lv, Jian and Cheng, Jun and Jia, Weile and Chen, Mohan and Ke, Guolin and E, Weinan and Zhang, Linfeng and Wang, Han},
  title = {{DPA-2: a large atomic model as a multi-task learner}},
  journal = {npj Comput Mater},
  year = {2024},
  volume = {10},
  pages = {293},
  doi = {10.1038/s41524-024-01493-2}
}

@article{pbMACEMP,
  author = {Batatia, Ilyes and Benner, Philipp and Chiang, Yuan and Elena, Alin M. and Kov{\'{a}}cs, D{\'{a}}vid P. and Riebesell, Janosh and Advincula, Xavier R. and Asta, Mark and Avaylon, Matthew and Baldwin, William J. and Berger, Fabian and Bernstein, Noam and Bhowmik, Arghya and Bigi, Filippo and Blau, Samuel M. and C{\u{a}}rare, Vlad and Ceriotti, Michele and Chong, Sanggyu and Darby, James P. and De, Sandip and Della Pia, Flaviano and Deringer, Volker L. and Elijo{\v{s}}ius, Rokas and El-Machachi, Zakariya and Fako, Edvin and Falcioni, Fabio and Ferrari, Andrea C. and Gardner, John L. A. and Gawkowski, Miko{\l}aj J. and Genreith-Schriever, Annalena and George, Janine and Goodall, Rhys E. A. and Grandel, Jonas and Grey, Clare P. and Grigorev, Petr and Han, Shuang and Handley, Will and Heenen, Hendrik H. and Hermansson, Kersti and Ho, Cheuk Hin and Hofmann, Stephan and Holm, Christian and Jaafar, Jad and Jakob, Konstantin S. and Jung, Hyunwook and Kapil, Venkat and Kaplan, Aaron D. and Karimitari, Nima and Kermode, James R. and Kourtis, Panagiotis and Kroupa, Namu and Kullgren, Jolla and Kuner, Matthew C. and Kuryla, Domantas and Liepuoniute, Guoda and Lin, Chen and Margraf, Johannes T. and Magd{\u{a}}u, Ioan-Bogdan and Michaelides, Angelos and Moore, J. Harry and Naik, Aakash A. and Niblett, Samuel P. and Norwood, Sam Walton and O'Neill, Niamh and Ortner, Christoph and Persson, Kristin A. and Reuter, Karsten and Rosen, Andrew S. and Rosset, Louise A. M. and Schaaf, Lars L. and Schran, Christoph and Shi, Benjamin X. and Sivonxay, Eric and Stenczel, Tam{\'{a}}s K. and Sutton, Christopher and Svahn, Viktor and Swinburne, Thomas D. and Tilly, Jules and van der Oord, Cas and Vargas, Santiago and Varga-Umbrich, Eszter and Vegge, Tejs and Vondr{\'{a}}k, Martin and Wang, Yangshuai and Witt, William C. and Wolf, Thomas and Zills, Fabian and Cs{\'{a}}nyi, G{\'{a}}bor},
  title = {{A foundation model for atomistic materials chemistry}},
  journal = {The Journal of Chemical Physics},
  year = {2025},
  volume = {163},
  pages = {184110},
  doi = {10.1063/5.0297006}
}

@article{pbChargE3,
  author = {Koker, Teddy and Quigley, Keegan and Taw, Eric and Tibbetts, Kevin and Li, Lin},
  title = {{Higher-order equivariant neural networks for charge density prediction in materials}},
  journal = {npj Comput Mater},
  year = {2024},
  volume = {10},
  pages = {161},
  doi = {10.1038/s41524-024-01343-1}
}

@article{pbEAC,
  author = {Qin, Xuejian and Lv, Taoyuze and Zhong, Zhicheng},
  title = {{EAC-Net: Predicting Real-Space Charge Density via Equivariant Atomic Contributions}},
  journal = {J. Chem. Theory Comput.},
  year = {2026},
  volume = {22},
  pages = {4813--4821},
  doi = {10.1021/acs.jctc.6c00283}
}

@article{pbEden,
  author = {Li, Xiwen and Xin, Zaizhou and Yu, Hongyu and Zhong, Yang and Hong, Liangliang and Gong, Xingao and Xiang, Hongjun},
  title = {{Efficient equivariant framework for universal charge density prediction}},
  journal = {Nat Mach Intell},
  year = {2026},
  volume = {8},
  pages = {1745--1755},
  doi = {10.1038/s42256-026-01326-5}
}

@article{pbOMat,
  author = {Barros-Luque, Luis and Shuaibi, Muhammed and Fu, Xiang and Wood, Brandon M. and Dzamba, Misko and Gao, Meng and Rizvi, Ammar and Uyttendaele, Matt and Zitnick, C. Lawrence and Ulissi, Zachary W.},
  title = {{The Open Materials 2024 (OMat24) inorganic materials dataset and models}},
  journal = {Nat Comput Sci},
  year = {2026},
  volume = {6},
  pages = {642--652},
  doi = {10.1038/s43588-026-00996-w}
}

@article{pbLewis,
  author = {Lewis, Alan M. and Grisafi, Andrea and Ceriotti, Michele and Rossi, Mariana},
  title = {{Learning Electron Densities in the Condensed Phase}},
  journal = {J. Chem. Theory Comput.},
  year = {2021},
  volume = {17},
  pages = {7203--7214},
  doi = {10.1021/acs.jctc.1c00576}
}

@article{pbMP,
  author = {Jain, Anubhav and Ong, Shyue Ping and Hautier, Geoffroy and Chen, Wei and Richards, William Davidson and Dacek, Stephen and Cholia, Shreyas and Gunter, Dan and Skinner, David and Ceder, Gerbrand and Persson, Kristin A.},
  title = {{Commentary: The Materials Project: A materials genome approach to accelerating materials innovation}},
  journal = {APL Materials},
  year = {2013},
  volume = {1},
  pages = {011002},
  doi = {10.1063/1.4812323}
}

@article{pbVASP,
  author = {Kresse, G. and Furthm{\"{u}}ller, J.},
  title = {{Efficient iterative schemes for
                    ab initio
                    total-energy calculations using a plane-wave basis set}},
  journal = {Phys. Rev. B},
  year = {1996},
  volume = {54},
  pages = {11169--11186},
  doi = {10.1103/PhysRevB.54.11169}
}

@article{pbPBE,
  author = {Perdew, John P. and Burke, Kieron and Ernzerhof, Matthias},
  title = {{Generalized Gradient Approximation Made Simple}},
  journal = {Phys. Rev. Lett.},
  year = {1996},
  volume = {77},
  pages = {3865--3868},
  doi = {10.1103/PhysRevLett.77.3865}
}

@article{pbPAW,
  author = {Bl{\"{o}}chl, P. E.},
  title = {{Projector augmented-wave method}},
  journal = {Phys. Rev. B},
  year = {1994},
  volume = {50},
  pages = {17953--17979},
  doi = {10.1103/PhysRevB.50.17953}
}

@article{pbTet,
  author = {Bl{\"{o}}chl, Peter E. and Jepsen, O. and Andersen, O. K.},
  title = {{Improved tetrahedron method for Brillouin-zone integrations}},
  journal = {Phys. Rev. B},
  year = {1994},
  volume = {49},
  pages = {16223--16233},
  doi = {10.1103/PhysRevB.49.16223}
}

@misc{pbMatter,
  author = {Yang, Han and Hu, Chenxi and Zhou, Yichi and Liu, Xixian and Shi, Yu and Li, Jielan and Li, Guanzhi and Chen, Zekun and Chen, Shuizhou and Zeni, Claudio and Horton, Matthew and Pinsler, Robert and Fowler, Andrew and Z{\"{u}}gner, Daniel and Xie, Tian and Smith, Jake and Sun, Lixin and Wang, Qian and Kong, Lingyu and Liu, Chang and Hao, Hongxia and Lu, Ziheng},
  title = {{MatterSim: A Deep Learning Atomistic Model Across Elements, Temperatures and Pressures}},
  year = {2024},
  eprint = {2405.04967},
  archivePrefix = {arXiv},
  url = {https://arxiv.org/abs/2405.04967}
}

@misc{pbXBridge,
  author = {Yang, Wooseong and Huang, Wei-Chieh and Zhang, Weizhi and Wang, Yu and Yu, Philip S. and Lee, Junhyun},
  title = {{XBridge: Entity-Grounded Latent Bridge for Heterogeneous LLM Communication}},
  year = {2026},
  eprint = {2608.11676},
  archivePrefix = {arXiv},
  url = {https://arxiv.org/abs/2608.11676}
}

@misc{pbLatBridge,
  author = {Zhu, Jiajun and Li, Haobo and Chen, Zixin and Qu, Huamin and Zhang, Wenshuo},
  title = {{LatBridge: Efficient and Stable Model Replacement in Latent-Communication Multi-Agent Systems}},
  year = {2026},
  eprint = {2610.27813},
  archivePrefix = {arXiv},
  url = {https://arxiv.org/abs/2610.27813}
}

@misc{pbLatentBridge,
  author = {Wang, Qichao and Qu, Tingyu and Wu, Pengcheng and Zhong, Yiran and Zhu, Feida and Shen, Zhiqi and Zhao, Peilin and Hoi, Steven},
  title = {{Latent Bridge: Asking for Help Without Words}},
  year = {2026},
  eprint = {2611.21301},
  archivePrefix = {arXiv},
  url = {https://arxiv.org/abs/2611.21301}
}

@misc{pbEqV3,
  author = {Liao, Yi-Lun and Hoffman, Alexander J. and Shen, Sabrina C. and Duval, Alexandre and Norwood, Sam Walton and Smidt, Tess},
  title = {{EquiformerV3: Scaling Efficient, Expressive, and General SE(3)-Equivariant Graph Attention Transformers}},
  year = {2026},
  eprint = {2604.09130},
  archivePrefix = {arXiv},
  note = {Appendix E},
  url = {https://arxiv.org/abs/2604.09130}
}

@inproceedings{pbStitch,
 author = {Bansal, Yamini and Nakkiran, Preetum and Barak, Boaz},
 booktitle = {Advances in Neural Information Processing Systems},
 editor = {M. Ranzato and A. Beygelzimer and Y. Dauphin and P.S. Liang and J. Wortman Vaughan},
 pages = {225--236},
 publisher = {Curran Associates, Inc.},
 title = {Revisiting Model Stitching to Compare Neural Representations},
 url = {https://proceedings.neurips.cc/paper_files/paper/2021/file/01ded4259d101feb739b06c399e9cd9c-Paper.pdf},
 volume = {34},
 year = {2021}
}

@inproceedings{pbMACEArch,
 author = {Batatia, Ilyes and Kovacs, David P and Simm, Gregor and Ortner, Christoph and Csanyi, Gabor},
 booktitle = {Advances in Neural Information Processing Systems},
 doi = {10.52202/068431-0830},
 editor = {S. Koyejo and S. Mohamed and A. Agarwal and D. Belgrave and K. Cho and A. Oh},
 pages = {11423--11436},
 publisher = {Curran Associates, Inc.},
 title = {MACE: Higher Order Equivariant Message Passing Neural Networks for Fast and Accurate Force Fields},
 url = {https://proceedings.neurips.cc/paper_files/paper/2022/file/4a36c3c51af11ed9f34615b81edb5bbc-Paper-Conference.pdf},
 volume = {35},
 year = {2022}
}

@misc{pbMACEOMAT,
  author = {{ACEsuit}},
  title = {{MACE-OMAT-0}},
  howpublished = {Model release},
  year = {2025},
  url = {https://github.com/ACEsuit/mace-foundations/releases/tag/mace_omat_0}
}

@misc{pbMACEMatPES,
  author = {{ACEsuit}},
  title = {{MACE-MATPES-0}},
  howpublished = {Model release},
  year = {2025},
  url = {https://github.com/ACEsuit/mace-foundations/releases/tag/mace_matpes_0}
}

@misc{pbMACEMH,
  author = {Batatia, Ilyes and Lin, Chen and Hart, Joseph and Kasoar, Elliott and Elena, Alin M. and Norwood, Sam Walton and Wolf, Thomas and Cs{\'{a}}nyi, G{\'{a}}bor},
  title = {{Cross Learning between Electronic Structure Theories for Unifying Molecular, Surface, and Inorganic Crystal Foundation Force Fields}},
  year = {2025},
  eprint = {2510.25380},
  archivePrefix = {arXiv},
  url = {https://arxiv.org/abs/2510.25380}
}

@misc{pbMagMACE,
  author = {Ho, Cheuk Hin and van der Oord, Cas and Darby, James P. and Keane, Theo and Benson, Raz L. and Espinoza, Cristian Rebolledo and Kulkarni, Rutvij and Spinu, Elina and Papanikolaou, Michail and Tomsett, Richard and Forrest, Robert M. and Bean, Jonathan J. and Cs{\'{a}}nyi, G{\'{a}}bor and Ortner, Christoph},
  title = {{Equivariant Many-body Message Passing Interatomic Potentials for Magnetic Materials}},
  year = {2026},
  eprint = {2604.08143},
  archivePrefix = {arXiv},
  url = {https://arxiv.org/abs/2604.08143}
}

@article{pbField,
  author = {Martin, Bradley A. A. and Ganose, Alex M. and Kapil, Venkat and Li, Tingwei and Butler, Keith T.},
  title = {{General Learning of the Electric Response of Inorganic Materials}},
  journal = {PRX Intelligence},
  year = {2026},
  volume = {1},
  pages = {013006},
  doi = {10.1103/b116-xy8k}
}

@misc{pbPolar,
  author = {Batatia, Ilyes and Baldwin, William J. and Kuryla, Domantas and Hart, Joseph and Kasoar, Elliott and Elena, Alin M. and Moore, Harry and Gawkowski, Mikołaj J. and Shi, Benjamin X. and Kapil, Venkat and Kourtis, Panagiotis and Magd{\u{a}}u, Ioan-Bogdan and Cs{\'{a}}nyi, G{\'{a}}bor},
  title = {{MACE-POLAR-1: A Polarisable Electrostatic Foundation Model for Molecular Chemistry}},
  year = {2026},
  eprint = {2602.19411},
  archivePrefix = {arXiv},
  url = {https://arxiv.org/abs/2602.19411}
}

@article{pbPETMAD,
  author = {Mazitov, Arslan and Bigi, Filippo and Kellner, Matthias and Pegolo, Paolo and Tisi, Davide and Fraux, Guillaume and Pozdnyakov, Sergey and Loche, Philip and Ceriotti, Michele},
  title = {{PET-MAD as a lightweight universal interatomic potential for advanced materials modeling}},
  journal = {Nat Commun},
  year = {2025},
  volume = {16},
  pages = {10653},
  doi = {10.1038/s41467-025-65662-7}
}

@article{pbGRACE,
  author = {Lysogorskiy, Yury and Bochkarev, Anton and Drautz, Ralf},
  title = {{Graph atomic cluster expansion for foundational machine learning interatomic potentials}},
  journal = {npj Comput Mater},
  year = {2026},
  volume = {12},
  pages = {114},
  doi = {10.1038/s41524-026-01979-1}
}

@article{pbNEP89,
  author = {Liang, Ting and Xu, Ke and Lindgren, Eric and Chen, Zherui and Zhao, Rui and Liu, Jiahui and Berger, Esm{\'{e}}e and Tang, Benrui and Zhang, Bohan and Wang, Yanzhou and Song, Keke and Ying, Penghua and Xu, Nan and Dong, Haikuan and Chen, Shunda and Erhart, Paul and Fan, Zheyong and Ala-Nissila, Tapio and Xu, Jianbin},
  title = {{NEP89: universal neuroevolution potential for inorganic and organic materials across 89 elements}},
  journal = {Nat Comput Sci},
  year = {2026},
  volume = {6},
  pages = {789--801},
  doi = {10.1038/s43588-026-01009-6}
}

@misc{pbEqV3Models,
  author = {{Mirror Physics}},
  title = {{EquiformerV3 model checkpoints}},
  howpublished = {Official model repository},
  year = {2026},
  url = {https://huggingface.co/mirror-physics/equiformer_v3}
}
\end{document}